
\input harvmac

 %
\catcode`@=11
\def\rlx{\relax\leavevmode}                  
 %
 %
 %
\font\tenmib=cmmib10
\font\sevenmib=cmmib10 at 7pt 
\font\fivemib=cmmib10 at 5pt  
\font\tenbsy=cmbsy10
\font\sevenbsy=cmbsy10 at 7pt 
\font\fivebsy=cmbsy10 at 5pt  
\def\BMfont{\textfont0\tenbf \scriptfont0\sevenbf
                              \scriptscriptfont0\fivebf
            \textfont1\tenmib \scriptfont1\sevenmib
                               \scriptscriptfont1\fivemib
            \textfont2\tenbsy \scriptfont2\sevenbsy
                               \scriptscriptfont2\fivebsy}
\def\BM#1{\rlx\ifmmode\mathchoice
                      {\hbox{$\BMfont#1$}}
                      {\hbox{$\BMfont#1$}}
                      {\hbox{$\scriptstyle\BMfont#1$}}
                      {\hbox{$\scriptscriptstyle\BMfont#1$}}
                 \else{$\BMfont#1$}\fi}
 %
 %
 %
 %
\def\inbar{\vrule height1.5ex width.4pt depth0pt}
\def\sinbar{\vrule height1ex width.35pt depth0pt}
\def\ssinbar{\vrule height.7ex width.3pt depth0pt}
\font\cmss=cmss10
\font\cmsss=cmss10 at 7pt
\def\ZZ{\rlx\leavevmode
             \ifmmode\mathchoice
                    {\hbox{\cmss Z\kern-.4em Z}}
                    {\hbox{\cmss Z\kern-.4em Z}}
                    {\lower.9pt\hbox{\cmsss Z\kern-.36em Z}}
                    {\lower1.2pt\hbox{\cmsss Z\kern-.36em Z}}
               \else{\cmss Z\kern-.4em Z}\fi}
\def\Ik{\rlx{\rm I\kern-.18em k}}  
\def\IC{\rlx\leavevmode
             \ifmmode\mathchoice
                    {\hbox{\kern.33em\inbar\kern-.3em{\rm C}}}
                    {\hbox{\kern.33em\inbar\kern-.3em{\rm C}}}
                    {\hbox{\kern.28em\sinbar\kern-.25em{\sevenrm C}}}
                    {\hbox{\kern.25em\ssinbar\kern-.22em{\fiverm C}}}
             \else{\hbox{\kern.3em\inbar\kern-.3em{\rm C}}}\fi}
\def\IP{\rlx{\rm I\kern-.18em P}}
\def\IR{\rlx{\rm I\kern-.18em R}}
\def\Ione{\rlx{\rm 1\kern-2.7pt l}}
 %
 %

 %

\def\intem#1{\par\leavevmode%
              \llap{\hbox to\parindent{\hss{#1}\hfill~}}\ignorespaces}
 %


 %
\newskip\humongous \humongous=0pt plus 1000pt minus 1000pt   
\def\caja{\mathsurround=0pt}
\newif\ifdtup
 %
\def\eqalign#1{\,\vcenter{\openup2\jot \caja
     \ialign{\strut \hfil$\displaystyle{##}$&$
      \displaystyle{{}##}$\hfil\crcr#1\crcr}}\,}
 %

 %

 %

 %

 %

 %

 %
 %
 %
 %
\def\,{\hskip1.5pt}           
 %
\let\a=\alpha

\let\c=\chi

\let\g=\gamma

\let\p=\pi

 %
 %
\def\Box{\sqcap\llap{$\sqcup$}}
\def\lapp{\lower.4ex\hbox{\rlap{$\sim$}} \raise.4ex\hbox{$<$}}
\def\gapp{\lower.4ex\hbox{\rlap{$\sim$}} \raise.4ex\hbox{$>$}}
\def\con{\ifmmode\raise.1ex\hbox{\bf*}
          \else\raise.1ex\hbox{\bf*}\fi}
\def\bo{{\raise.15ex\hbox{\large$\Box\kern-.39em$}}}

\def\dual{\relax\leavevmode\lower.9ex\hbox{\titlerms*}}

\let\8=\otimes
 %
 %
 %
 %

\let\2=\underline

 %
\def\dt#1{{\buildrel{\smash{\lower1pt\hbox{.}}}\over{#1}}}

\font\eightrm=cmr8
\def\6(#1){\relax\leavevmode\hbox{\eightrm(}#1\hbox{\eightrm)}}
\def\0#1{\relax\ifmmode\mathaccent"7017{#1}     
                \else\accent23#1\relax\fi}      
\def\7#1#2{{\mathop{\null#2}\limits^{#1}}}      
\def\5#1#2{{\mathop{\null#2}\limits_{#1}}}      
 %

 %

 %

 %

 %
\newbox\t@b@x
\def\rightarrowfill{$\m@th \mathord- \mkern-6mu
     \cleaders\hbox{$\mkern-2mu \mathord- \mkern-2mu$}\hfill
      \mkern-6mu \mathord\rightarrow$}
\def\tooo#1{\setbox\t@b@x=\hbox{$\scriptstyle#1$}%
             \mathrel{\mathop{\hbox to\wd\t@b@x{\rightarrowfill}}%
              \limits^{#1}}\,}
\def\leftarrowfill{$\m@th \mathord\leftarrow \mkern-6mu
     \cleaders\hbox{$\mkern-2mu \mathord- \mkern-2mu$}\hfill
      \mkern-6mu \mathord-$}
\def\froo#1{\setbox\t@b@x=\hbox{$\scriptstyle#1$}%
             \mathrel{\mathop{\hbox to\wd\t@b@x{\leftarrowfill}}%
              \limits^{#1}}\,}
 %
\def\frac#1#2{{#1\over#2}}
\def\frc#1#2{\relax\ifmmode{\textstyle{#1\over#2}} 
                    \else$#1\over#2$\fi}           
 %
\def\Claim#1#2#3{\bigskip\begingroup%
                  \xdef #1{\secsym\the\meqno}%
                   \writedef{#1\leftbracket#1}%
                    \global\advance\meqno by1\wrlabeL#1%
                     \noindent{\bf#2}\,#1{}\,:~\sl#3\vskip1mm\endgroup}

\def\QED{\rlx\hfill$\Box$\kern-7pt\raise3pt\hbox{$\surd$}\bigskip}
 %
 %

 %
\def\muthstrut{\vphantom1}
\def\mutrix#1{\null\,\vcenter{\normalbaselines\m@th
        \ialign{\hfil$##$\hfil&&~\hfil$##$\hfill\crcr
            \muthstrut\crcr\noalign{\kern-\baselineskip}
            #1\crcr\muthstrut\crcr\noalign{\kern-\baselineskip}}}\,}

 %
\def\YT#1#2{\vcenter{\hbox{\vbox{\baselineskip0pt\parskip=\medskipamount%
             \def\Box{$\sqcap\llap{$\sqcup$}$\kern-1.2pt}%
              \def\Z{\hfil\vskip-5.8pt}\lineskiplimit0pt\lineskip0pt%
               \setbox0=\hbox{#1}\hsize\wd0\parindent=0pt#2}\,}}}
\def\EU{\rlx\ifmmode \c_{{}_E} \else$\c_{{}_E}$\fi}
\def\TM{\rlx\ifmmode {\cal T_M} \else$\cal T_M$\fi}
\def\TW{\rlx\ifmmode {\cal T_W} \else$\cal T_W$\fi}
\def\CM{\rlx\ifmmode {\cal T\rlap{\bf*}\!\!_M}
             \else$\cal T\rlap{\bf*}\!\!_M$\fi}
\def\hm#1#2{\rlx\ifmmode H^{#1}({\cal M},{#2})
                 \else$H^{#1}({\cal M},{#2})$\fi}
\def\CP#1{\rlx\ifmmode\IP^{#1}\else\IP$^{#1}$\fi}
\def\cP#1{\rlx\ifmmode\IC{\rm P}^{#1}\else$\IC{\rm P}^{#1}$\fi}

\def\sll#1{\rlx\rlap{\,\raise1pt\hbox{/}}{#1}}
\def\Sll#1{\rlx\rlap{\,\kern.6pt\raise1pt\hbox{/}}{#1}\kern-.6pt}
%

 %
 %
\def\ie{\hbox{\it i.e.}}        

\def\CY{Calabi-\kern-.2em Yau}

\def\3{\ifmmode\ldots\else$\ldots$\fi}
\def\Z{\hfil\break\rlx\hbox{}\quad}
\def\3{\ifmmode\ldots\else$\ldots$\fi}
\def\?{d\kern-.3em\raise.64ex\hbox{-}}           
\def\9{\raise.43ex\hbox{-}\kern-.37em D}         

 %
 %

 %

 %

\def\NP#1{{\it Nucl.\,Phys.\,}{\bf#1\,}}
\def\PL#1{{\it Phys.\,Lett.\,}{\bf#1\,}}

\def\MPL#1{{\it Mod.\,Phys.\,Lett.\,}{\bf#1\,}}
\def\PRL#1{{\it Phys.\,Rev.\,Lett.\,}{\bf#1\,}}
\def\CMP#1{{\it Commun.\,Math.\,Phys.\,}{\bf#1\,}}

 %
 %
 %
\baselineskip=13.0861pt plus2pt minus1pt
\parskip=\medskipamount
\let\ft=\foot
\noblackbox
\def\SaveTimber{\abovedisplayskip=1.5ex plus.3ex minus.5ex
                \belowdisplayskip=1.5ex plus.3ex minus.5ex
                \abovedisplayshortskip=.2ex plus.2ex minus.4ex
                \belowdisplayshortskip=1.5ex plus.2ex minus.4ex
                \baselineskip=12pt plus1pt minus.5pt
 \parskip=\smallskipamount
 \def\ft##1{\unskip\,\begingroup\footskip9pt plus1pt minus1pt\setbox%
             \strutbox=\hbox{\vrule height6pt depth4.5pt width0pt}%
              \global\advance\ftno by1\footnote{$^{\the\ftno)}$}{##1}%
               \endgroup}
 \def\listrefs{\footatend\vfill\immediate\closeout\rfile%
                \writestoppt\baselineskip=10pt%
                 \centerline{{\bf References}}%
                  \bigskip{\frenchspacing\parindent=20pt\escapechar=` %
                   \rightskip=0pt plus4em\spaceskip=.3333em%
                    \input refs.tmp\vfill\eject}\nonfrenchspacing}}
 %
\def\Afour{\ifx\answ\bigans
            \hsize=16.5truecm\vsize=24.7truecm
             \else
              \hsize=24.7truecm\vsize=16.5truecm
               \fi}
\catcode`@=12

\let\al=\alpha

\def\eqaligntwo#1{\,\vcenter{\openup2\jot \caja
     \ialign{\strut \hfil$\displaystyle{##}$&
                         $\displaystyle{{}##}$\hfil&
                          $\displaystyle{{}##}$\hfil\crcr#1\crcr}}\,}

 \def\Afour{\hsize=16.5truecm\vsize=24.7truecm}

\baselineskip=12pt

\Title{\vbox{\baselineskip12pt \hbox{IASSNS-HEP-94/11}}}
       {\vbox{\centerline{On the Elliptic Genus and
        Mirror Symmetry}}}

\centerline{Per Berglund\footnote{$^{\dag}$}
{Email: berglund@guinness.ias.edu}
 and M{\aa}ns Henningson\footnote{$^{\ddag}$}
{Email: mans@guinness.ias.edu}} \vskip4mm
 \centerline{\it School of Natural Science} \vskip 0mm
 \centerline{\it Institute for Advanced Study}      \vskip 0mm
 \centerline{\it Olden Lane}               \vskip 0mm
 \centerline{\it Princeton, NJ 08540}                \vskip 0mm
 \vfill

\centerline{ABSTRACT}\vskip5mm
\vbox{
\baselineskip=12pt\noindent
The elliptic genus for arbitrary
two dimensional $N=2$ Landau-Ginzburg orbifolds is computed. This is
used to search for possible mirror pairs of such models.
An important aspect of this work is that there is no restriction to
theories for which the conformal anomaly is $\hat c\in\ZZ$,
 nor are the
results only valid at the conformal fixed point.
}

\vfill
\centerline{{To appear in {\it Essays on Mirror Manifolds II}}}

\Date{\vbox{
       \line{1/94 \hfill}}}

\noblackbox
\lref\rDixon{For a review and references, see L.~Dixon, in {\it
     Superstrings, Unified Theories and Cosmology 1987},
     eds.~G.~Furlan et al.\ (World Scientific, Singapore, 1988).}

\lref\rCHSW{P.~Candelas, G.~Horowitz, A.~Strominger and E.~Witten,
      \NP{B258} (1985) 46.}

\lref\rMPR{P.~Candelas, M.~Lynker and R.~Schimmrigk,
      \NP{B341} (1990) 383\semi
      M.~Kreuzer and H.~Skarke, \CMP{150} (1992) 137\semi
      A.~Klemm and R.~Schimmrigk~, ``Landau-Ginzburg String Vacua'',
       {\it CERN preprint} CERN-TH 6459/92\semi
          M.~Kreuzer and H.~Skarke, \NP{B388} (1992) 113.}
\lref\rMax{M.~Kreuzer and H.~Skarke, ``All Abelian Symmetries of
       Landau-Ginzburg Potentials'', CERN-TH-6705-92.}

\lref\rGP{B.~R.~Greene and M.~R.~Plesser, \NP{B338} (1990) 15.}

\lref\rVW{C.~Vafa and N.P.~Warner, \PL{218B} (1989) 377\semi
          E.Martinec, \PL{B217} (1989) 431.}

\lref\rRolf{R.~Schimmrigk, \PRL{70} (1993) 3688.}

\lref\rBH{P.~Berglund and T.~H\"ubsch, in {\it Essays on Mirror Manifolds},
S.T.~Yau ed. (International Press, Hong Kong 1992); \NP{B393} (1993) 377.}

\lref\rIV{C.~Vafa, \MPL{A4} (1989) 1169\semi
          K.~Intrilligator and C.~Vafa, \NP{B339} (1990) 95.}

\lref\rASY{O.~Aharony, A.N.~Schellekens and S.~Yankielowicz,
``Charge Sum Rules in $N=2$ Theories'', NIKHEF-H/93-27 and
Tel-Aviv University report TAUP-2123-93.}

\lref\rFY{P.~Di~Francesco and S.~Yankielowicz,
\NP{B409} (1993) 186.}

\lref\rMH{M.~Henningson, ``N=2 Gauged WZW models and the Elliptic Genus'',
IAS preprint IASSNS-HEP-39/93, \NP{B} to appear.}

\lref\rAFY{O.~Aharony, P.~Di~Francesco and S.~Yankielowicz,
``Elliptic Genera and the Landau-Ginzburg Approach to $N=2$ Orbifolds'',
Saclay preprint SPhT 93/068 and Tel-Aviv University report TAUP 2069-93\semi
           T.~Kawai, Y.~Yamada and S.-K.~Yang,
``Elliptic Genera and $N=2$ Superconformal Field Theory'',
KEK-TH-362.}

\lref\rWitten{E.~Witten, ``On the Landau-Ginzburg Description of $N=2$
minimal models'', IAS preprint IASSNS-HEP-93/10.}

\lref\rLVW{W.~Lerche, C.~Vafa and N.P.~Warner, \NP{B324} (1989) 427.}

\lref\rVT{C.~Vafa, \NP{B273} (1986) 592.}

\lref\rGQ{D.~Gepner and Z.~Qiu, \NP{B285} (1987) 423.}

\lref\rSW{A. Schellekens and N. Warner, \PL{177B} (1986) 317\semi
A. Schellekens and N. Warner, \PL{181B} (1986) 339\semi
K. Pilch, A. Schellekens and N. Warner, \NP{B287} (1987) 317.}

\lref\rWittenCMP{E. Witten, \CMP{\bf 109} (1987) 525 \semi
E. Witten, in {\it Elliptic Curves and Modular Forms in Algebraic Topology} ed.
P. Landwebber, (Springer Verlag 1988).}

\lref\rGSW{M.B. Green, J.H. Schwarz and E. Witten, {\it Superstring
Theory} (Cambridge monographs on mathematical physics 1987).}

\lref\rWW{E.T.~Whittaker and G.N.~Watson, {\it A Course of Modern Analysis},
(Cambridge University Press 1958).}

\lref\rBatdual{V.~Batyrev, ``Dual Polyhedra and Mirror Symmetry for Calabi-Yau
Hypersurfaces in Toric Varieties'', Preprint, November 18, 1992\semi
      P.~Candelas, X.~de~la~Ossa and S.~Katz, ``Mirror Symmetry for
Calabi-Yau Hypersurfaces in Weighted $\IP^4$ and an Extension of
Landau-Ginzburg Theory'', in preparation.}

\lref\rAGM{P.S.~Aspinwall, B.R.~Greene and D.R.~Morrison, ``\CY\ Moduli
      Space, Mirror Manifolds and Spacetime Topology Change'', IAS preprint
      IASSNS-HEP-93/38.}

\lref\rArnold{V.I.~Arnold, S.M.~Gusein-Zade and A.N.~Varchenko, {\it
     Singularities of Differentiable Maps, Vol.~I}~ (Birkh\"auser,
     Boston, 1985).}

\lref\rCdGP{P.~Candelas, X.~ de la Ossa, P.~Green and L.~Parkes,
       \NP{B359}~(1991)~21.}

\lref\rDHVW{L.~Dixon, J.~Harvey, C.~Vafa and E.~Witten, \NP{B261} (1985)
678\semi
L.~Dixon, J.~Harvey, C.~Vafa and E.~Witten, \NP{B274} (1986) 285.}

\lref\rKS{Y.~Kazama and H.~Suzuki, \NP{B321} (1989) 232\semi
Y.Kazama and H. Suzuki, \PL{216B}~(1989)~112.}

\lref\rPM{P.~Berglund and M.~Henningson, ``Landau-Ginzburg Orbfolds,
Mirror Symmetry and the Elliptic Genus'', IAS preprint IASSNS-HEP-93/92,
submitted to \NP{B}.}

\lref\rMM{{\it Essays on Mirror Manifolds},
S.T.~Yau ed. (International Press, Hong Kong 1992).}

\newsec{Introduction}\noindent
$N=2$ superconformal field  theories have attracted much attention during
the last couple of years in particular because of their importance in
understanding the  compactification of
the heterotic string. Although $(0,2)$ models are sufficient for
$N=1$ space-time supersymmetry in the effective four dimensional  low-energy
theory most effort has been concentrated on the left-right symmetric $(2,2)$
theories~\rDixon.
One example of such theories, which is the focus of the present article,
is the class of $N=2$ Landau-Ginzburg vacua. Though not conformally invariant
as they stand, they are believed to flow to a fixed point in the infrared,
characterized by the superpotential $W$, a quasi homogeneous function
of the complex chiral superfields~\rVW.
Assuming that the original theory is invariant under
a symmetry group $H$, one can  consider a new set of models
through the process of orbifolding~\rDHVW.
Field configurations in the orbifold model are then identified
modulo the action of the group $H$. The process of orbifolding will
turn out to be an integral part of the construction of mirror pairs.

Previous studies of $(2,2)$ compactifications in general and $N=2$
Landau-Ginzburg models (including orbifold constructions) in
particular~\refs{\rGP,\rMPR,\rBH,\rCdGP,\rMax,\rBatdual,\rAGM}
have indicated that there exists a symmetry relating seemingly
different models; mirror
symmetry~\ft{For a review of mirror symmetry, see other articles in this
volume as well as~\rMM.}.
At the level of conformal field theory the symmetry can be
formulated as an isomorphism
between two theories, the only difference being a change of the sign of the
left-moving $U(1)$-generator. As trivial as it may seem it has far
far reaching consequences. For example, in terms of \CY\ sigma
models the effect of
mirror symmetry is to equate the space-time physics of target spaces of
different topology.

Unfortunately, the original construction by Greene and Plesser~\rGP,
who considered orbifolds of tensor products of
$N=2$ minimal models,
is still the only one for which mirror
symmetry has been rigorously proven at the level of conformal field
theory. The problem is that in order to compare the theories we have
to compute the partition function $Z(q,\gamma_L,\gamma_R) = {\rm Tr}
(-1)^F q^{L_0} {\bar q}^{\bar L_0} \exp (i\gamma_L J_0 + i\gamma_R {\bar
J}_0)$ where $L_0$ ($\bar{L}_0$) and $J_0$ (${\bar J}_0$) are the energy
and $U(1)$
charge operator respectively of the left-moving (right-moving) $N=2$ algebra.
Apart from the minimal models and their orbifolds this is
 in general not a feasible task  for an $N=2$ theory.
The situation is much better for the elliptic genus~\refs{\rSW\,\rWittenCMP},
which is simply the
restriction of the partition function to $\gamma_R=0$, \ie\ $Z(q,\gamma,0)$.
This restricted partition function
  has an interpretation as an index of the $N=1$ right-moving
supercharge. This property was
recently used by Witten~\rWitten\ to calculate the elliptic genera of certain
Landau-Ginzburg models which are believed to flow to the minimal models. The
results were compared to the elliptic genera of the minimal models, calculated
from the known characters of the $N=2$ discrete series representations,
in~\refs{\rWitten,\rFY}. The affirmative outcome of these computations lends
further support to the conjectured isomorphism between Landau-Ginzburg
theories (at the conformal fixed point) and the minimal series~\rLVW.
For other applications of the elliptic genus,
see~\refs{\rMH,\rAFY,\rASY}.

In the following we will use the elliptic genus as a tool for studying
mirror symmetry in the context of Landau-Ginzburg orbifolds. Since the
spectrum of an $N=2$ theory is symmetric under charge conjugation, the elliptic
genus is invariant under $J_0 \rightarrow -J_0$, \ie\ $Z(q, \gamma, 0) =  Z(q,
-\gamma, 0)$. Thus, the
 elliptic genera of two models that constitute a mirror pair
must therefore be equal (up to a sign, which could be thought of as arising
from different normalizations of the path integral measures of the two
theories).
Although the equality of the elliptic genera is
only an indication of mirror symmetry, and by no means a
proof, we will for brevity
refer to two models with the same elliptic genus as constituting a mirror pair.

Finally, we want to stress a point which is not of direct relevance for
our work but which may have its own merits. Mirror symmetry is mostly
discussed in the context of models with $\hat c=3$ (or at least
with integer $\hat c$). The theory may then in many cases
be interpreted as a sigma model where the target space is a (complex)
$\hat c$ dimensional \CY.
(For a discussion of theories which may have a geometric interpretation
in a generalized sense, see for example~\rRolf.)
However, mirror symmetry seems to be an inherent two dimensional feature
and the value of $\hat c$ is of less importance. In addition,
the conformal aspect could also turn out to be of subordinate value
since our computations  indicate that mirror symmetry may be a
property of any two dimensional $N=2$ quantum field theory.
In all this is telling us to study mirror symmetry in a broader context
and hopefully this will teach us about phenomena relevant to string vacua
as well.

This article is organized as follows: In section~2, we calculate the elliptic
genus for an arbitrary Landau-Ginzburg orbifold as well as the
Poincar\'{e} polynomial of the theory by taking the $q \rightarrow 0$ limit
of the elliptic genus.
In section~3, we
discuss a plausible scenario and find a sufficient condition
for mirror symmetry between Landau-Ginzburg
orbifolds.
Indeed, it can be shown~\rPM\ that the models proposed in~\rBH\ satisfy this
condition. We conjecture that all pairs of conjugate Landau-Ginzburg models may
be obtained by taking products of these models.

\newsec{Landau-Ginzburg orbifolds and the elliptic genus}
\subsec{The elliptic genus}\noindent
Let us start by reviewing some important facts about $N=2$
Landau-Ginzburg models and their orbifolds.
Consider the action of a $(2,2)$ Landau-Ginzburg model written in superspace as
\eqn\eaction{
S = \int d^2 z \, d^4 \theta \, K(X_i, \bar{X}_i) + \epsilon \int d^2 z \, d^2
\theta \, W(X_i) + {\rm c.c.}.
}
The $X_i$ for $i \in N$, with the total number of fields denoted by $|N|$,
 are complex chiral superfields with component
expansions
\eqn\esuperf{
X_i = x_i + \theta_+ \psi_i^+ + \theta_- \psi_i^- + \theta_+ \theta_- F_i.
}
The superpotential $W$ is a holomorphic and
quasi-homogeneous function of the $X_i$, \ie\ it should be possible to assign
some weights $k_i \in \ZZ$ to the fields $X_i$ for $i \in N$ and a degree
of homogeneity $D \in \ZZ$ to $W$ such that
\eqn\escale{
W(\lambda^{k_i} X_i) = \lambda^D W(X_i).
}
The central charge for the theory is given by $\hat c=\sum_i(1-2q_i)$
with $q_i=k_i/d$ the left-(and right-) moving $U(1)$-charge~\rVW.
The model defined by~\eaction\ is believed to flow to a conformally
invariant model in the infrared. Under this renormalization group flow, the
K\"ahler potential $K(X_i, \bar{X}_i)$ will get renormalized in some
complicated way, but there are strong reasons to believe that the
superpotential $W(X_i)$ is an invariant of the flow~\rVW.

In general, $W$ will be invariant under a discrete, abelian group $G$ of phase
symmetries. The fields $X_i$ transform in some representations $R_i$ under $G$.
In the following, we denote the set of representations $\{ R_i \}$ for $i \in
N$ collectively as $R$. A representation $R_i$ of $G$ is specified by a
function $R_i(g) = \exp (i 2 \pi \theta_i(g))$ defined for $g \in G$, which
fulfills $R_i(g_1 g_2) = R_i(g_1) R_i(g_2)$ for all $g_1, g_2 \in G$. The
invariance of $W$ means that
\eqn\eXXX{
W(R_i(g) X_i) = W(X_i) \;\; {\rm for} \;\; g \in G.
}
By taking $\lambda = \exp i 2 \pi/D$ in~\escale, we see that $G$ will
always contain an element $q$ such that $R_i(q) = \exp i 2 \pi q_i$ for $i \in
N$. From a theory invariant under some symmetry group $H$, we may construct a
new theory by taking the $H$ orbifold of the original theory, \ie\ by modding
out by the action of $H$. In our case, $H$ could be any subgroup of the group
$G$ of phase symmetries of $W$. We will denote the theory thus obtained as
$W/H$~\ft{In general $q\notin H$ and hence $W/H$ is not a valid superstring
vacuum, even if $\hat c = 3$.}.

Our object is to calculate the elliptic genus of $W/H$. This calculation is
feasible because of the invariance of the elliptic genus under deformations of
the theory which preserve the right-moving supersymmetry~\rWitten.
We may therefore
smoothly turn off the superpotential interactions by letting $\epsilon
\rightarrow 0$, which turns the model into a free field theory. (We take $K$ to
be the K\"ahler potential of $\IC^{|N|}$ with the flat metric.)
The elliptic genus of the model is
determined by the set of representations $R$ and the group $H$ and may be
denoted as $Z[R/H]$, where we have suppressed the dependence on $\gamma$ and
$q$. Being essentially a genus one correlation function, it may be written as
\eqn\epartition{
Z[R/H] = \frac{1}{|H|} \sum_{h_a, h_b \in H} Z[R](h_a, h_b).
}
Here $Z[R](h_a, h_b)$ denotes the contribution from field configurations
twisted by $h_a$ and $h_b$ around the $a$ and $b$ cycles of the torus
respectively, and $|H|$ is the number of elements of $H$.
In the free field
limit, the contribution from each twist sector is a product of contributions
from each of the fields $X_i$ in the theory, \ie\
\eqn\eZsector{
Z[R](h_a, h_b) = \prod_{i \in N} Z[R_i](h_a, h_b).
}
A by now standard
calculation gives~\refs{\rWitten,\rPM}
\eqn\ethreesix{
Z[R_i](h_a, h_b) = e^{-i 2 \pi \gamma \theta_i(h_a)} \frac{\Theta_1((1-q_i)
\gamma - \theta_i(h_b) - \tau \theta_i(h_a) | \tau)}{\Theta_1(q_i \gamma +
\theta_i (h_b) + \tau \theta_i (h_a) | \tau)}.
}
where $\Theta_1$ is the Jacobi $\Theta_1$ function~\rGSW.
The prefactor is chosen such that the elliptic genus is well-defined,
\ie\ it should only depend on $R_i(h_a)=exp(i2\p \theta_i(h_a))$ and
$R_i(h_b)$.
Furthermore, with this choice the complete elliptic genus,
given by~\epartition\ with~\eZsector, has the correct
double quasi periodicity in $\g$ and modular
transformation properties~\rPM~\ft{Although the canonical choice,
nontrivial phase factors can be introduced, so called discrete torsion~\rVT.}.

\subsec{The Poincar\'{e} polynomial}
\noindent
A necessary condition for the elliptic genera of two $N = 2$ models to be equal
is that they coincide in the $\tau \rightarrow i \infty$ limit, \ie\ that the
two models have the same Poincar\'{e} polynomial. Following
Francesco and Yankielowicz~\rFY, this condition can be shown  to be
sufficient in the
case of orbifolds of Landau-Ginzburg models with isomorphic groups of phase
symmetries~\rPM~\ft{In a more general case it may not be sufficient to merely
consider the $q\to 0$ limit~\rASY.}.

The Poincar\'{e} polynomial is a polynomial in $t^{1/D}$, where $t = \exp (i 2
\pi
\gamma)$ and $D$ is some integer (defined by~\escale\ for the case of
Landau-Ginzburg orbifolds). It can be seen as the generating function for the
charges of the Ramond sector ground states under the $U(1)$ symmetry of the
left-moving $N=2$ algebra. From our results in the last section,
the Poincar\'{e} polynomial for a Landau-Ginzburg orbifold may be written as a
sum over contributions from different twist-sectors;
\eqn\efourthree{
P[R/H] = \frac{1}{|H|} \sum_{h_a, h_b \in H} P[R](h_a, h_b).
}
Furthermore, the contribution from each twist-sector is a product of the
contributions from each of the fields;
\eqn\efourfour{
P[R](h_a, h_b) = \prod_{i \in N} P[R_i](h_a, h_b).
}
Finally, by taking the $\tau \rightarrow i \infty$ limit of~\ethreesix, we
find that
\eqn\efourfive{
P[R_i](h_a, h_b) = \left\{
\eqaligntwo{
&- t^{1/2 {-} [\theta_i(h_a)]} \qquad & {\rm for} \; R_i(h_a) \neq 1 \cr
&t^{-1/2} (t^{q_i} R_i(h_b) {-} t)(1 {-} t^{q_i} R_i(h_b))^{-1}
\qquad & {\rm for} \; R_i(h_a) = 1 } \right. .
}
Here, $[x]$ denotes the fractional part of $x$, \ie\ $[x] = x {\;\; \rm mod
\;\;} \ZZ$ and $0 \leq [x] < 1$. Note that if $R_i(h_a) = 1$, or equivalently
$[\theta_i(h_a)] = 0$, then the field $X_i$ is left untwisted by the
transformation $h_a$. It will prove convenient to introduce
\eqn\eptw{
P^{\rm tw}[R](g) = \prod_{i \in N} (- t^{1/2})^{{\rm tw}[R](g)} \, t^{-\sum_{i
\in N} [\theta_i(g)]}
}
and
\eqn\epinv{
P^{\rm inv}[R_i](g) = t^{-1/2} (t^{q_i} R_i(g) - t)(1 - t^{q_i}
R_i(g))^{-1},
}
where ${\rm tw}[R](g)$ denotes the number of fields that are twisted by $g$. We
may then write
\eqn\efourtw{
P[R](h_a, h_b) = P^{\rm tw}[R](h_a) \, \prod_{R_i(h_a) = 1} P^{\rm
inv}[R_i](h_b).
}

Incidentally, there is a generalized Poincar\'{e} polynomial, which is
sensitive to the charges of the Ramond sector ground states under both the
left- and right-moving $U(1)$ symmetry. Since mirror symmetry acts by reversing
the sign of one of the $U(1)$ charges
and leaving the other unaffected, this generalized Poincar\'{e} polynomial is
useful to check whether two models might be each others mirror partners rather
than being completely equivalent theories. It may be defined as~\rLVW\
\eqn\eXXX{
P(t, \bar{t}) = {\rm Tr} (t^{J_0} \bar{t}^{\bar{J}_0}),
}
where the trace is over the ground states in the Ramond sector. For a
Landau-Ginzburg orbifold, this generalized Poincar\'{e} polynomial has been
calculated by Intrilligator and Vafa~\refs{\rIV}. The only difference with
respect to the Poincar\'{e} polynomial that we have discussed is that the
arguments of $P^{\rm tw}[R](g_a)$ and $P^{\rm inv}[R_i](g_b)$ in~\efourtw\ are
$t/\bar{t}$ and $t\bar{t}$ respectively instead of $t$. We may therefore
continue to work with the restricted Poincar\'{e} polynomial which only depends
on $t$. A criterion for mirror symmetry is then that the Poincar\'{e}
polynomials of the two models must be equal and that contributions from twisted
(untwisted) fields in one model should correspond to contributions from
untwisted (twisted) fields in the other.

\newsec{Mirror symmetry for Landau-Ginzburg orbifolds}
\subsec{General results}\noindent
Suppose that we have two Landau-Ginzburg models, each with  $|N|$
superfields, the phase symmetry groups of
which are both isomorphic to the same abelian group $G$. We denote the sets of
representations as $R$ and $\tilde{R}$ respectively for the two models. In
general, we distinguish all quantities pertaining to the second model with a
tilde. We are interested in the situation in which the $H$ orbifold of the
first model is the mirror partner of the $\tilde{H}$ orbifold of the second
model for some subgroups $H$ and $\tilde{H}$ of $G$. This means that their
elliptic genera   should be equal up to a sign:
\eqn\enfivezero{
Z[R/H] = \pm Z[\tilde{R} / \tilde{H}].
}
We will propose a
natural way for two models to be each others mirror partners in this sense,
but to do so, we first need to discuss some aspects of abelian groups.

All irreducible representations of the abelian group $G$ are one-dimensional.
Given two irreducible representations we may construct a new irreducible
representation by taking their tensor product. Clearly, the set of irreducible
representations of $G$ form a group $G^*$ under the tensor product $\otimes$.
This group is in fact isomorphic to $G$ itself. Given a subgroup $H$ of $G$ we
define its dual as the subgroup $\tilde{H}$ of $G^*$ of representations on
which $H$ is trivially represented, \ie\ $\tilde{H}$ is the set of $R \in G^*$
such that $R(g) = 1$ for $g \in H$. In particular, the dual of $G$ itself is
the trivial subgroup of $G^*$ which only consists of the identity element, and
the dual of the trivial subgroup of $G$ is $G^*$. Clearly, if $H_1 \subset H_2$
then $\tilde{H_2} \subset \tilde{H_1}$. Similarly, we note that a
representation of the group $G^*$ has a natural interpretation as an element of
$G$. Therefore, given a subgroup $\tilde{H}$ of $G^*$ we may define its dual
$\tilde{\tilde{H}}$ as
 the subgroup of elements of $G$ which are trivially represented by all $R \in
\tilde{H}$. We see that $\tilde{\tilde{H}} = H$. The situation is thus
completely symmetric under interchange of $G$ and $G^*$.
As an alternative definition of the duality between $H$ and $\tilde H$
we have the following relation
\eqn\enfiveone{
\frac{1}{|H|} \sum_{g \in H} \tilde{g}(g) = \left\{ \eqalign{ 1 & \quad{\rm
for \;\;} \tilde{g} \in \tilde{H} \cr 0 & \quad{\rm otherwise} } \right.
}
and its partner obtained by changing the roles of $G$ and $G^*$. The
summand $\tilde g(g)$ is the function, defined on $G$, which
specifies the $G$ representation $\tilde g\in G^*$.

We now interpret our candidate mirror pair of Landau-Ginzburg models so that
the fields of the first model transform in the representations $R_i$ for $i \in
N$ under the symmetry group $G$, whereas the fields of the second model
transform in the representations $\tilde{R}_i$ for
$i \in \tilde{N}$ under the group $G^*$. As our notation suggests,
we will take the $H$ orbifold of the first model and the $\tilde{H}$ orbifold
of the second model, where $H$ and $\tilde{H}$ are dual subgroups.

For $g \in
G$ and $\tilde{g} \in G^*$ we now
define the (partial) Fourier transform of the
elliptic genus contributions as
\eqn\eXXX{
\hat{Z}[R](g, \tilde{g}) = \frac{1}{|G|} \sum_{g^\prime \in G} \tilde{g}
(g^\prime) Z[R](g, g^\prime),
}
which may be inverted by means of~\enfiveone\ to yield
\eqn\eXXX{
Z[R](g, g^\prime) = \sum_{\tilde{g} \in G^*} \tilde{g}^{-1}(g^\prime)
\hat{Z}[R](g,
\tilde{g}).
}
Inserting this in~\epartition\ and using~\enfiveone\ we get
\eqn\eXXX{
Z[R/H] = \sum_{h \in H} \sum_{\tilde{h} \in \tilde{H}} \hat{Z}[R](h,
\tilde{h}).
}
Analogously, we calculate
\eqn\eXXX{
Z[\tilde{R}/\tilde{H}] = \sum_{\tilde{h} \in \tilde{H}} \sum_{h \in H}
\hat{Z}[\tilde{R}](\tilde{h}, h).
}
A very natural way to fulfill~\enfivezero\ is then to require that
\eqn\enfivetwo{
\hat{Z}[R](g, \tilde{g}) = \pm \hat{Z}[\tilde{R}](\tilde{g}, g)
}
for $g \in G$ and $\tilde{g} \in G^*$.
We will say that  sets of representations $R$ and $\tilde R$ which
fulfill~\enfivetwo\ are conjugates of each other.
Note that this condition implies
\enfivezero\ for {\it any} $H$ and its dual $\tilde{H}$. Furthermore, given two
Landau-Ginzburg models $R_1$ and $R_2$ with symmetry groups isomorphic to $G_1$
and $G_2$ respectively, we may construct the product model $R = R_1 \times R_2$
with symmetry group $G \simeq G_1 \times G_2$. If now $R_1$ and $R_2$ are
conjugates to $\tilde{R}_1$ and $\tilde{R}_2$ respectively so that each
 pair satisfies~\enfivetwo, then the pair of product models $R = R_1
\times R_2$ and $\tilde{R} = \tilde{R}_1 \times \tilde{R}_2$ also satisfies
\enfivetwo. This means that any $H$ orbifold of $R$, for $H$ a subgroup of
$G \simeq G_1 \times G_2$, will be the mirror partner of the corresponding
$\tilde{H}$ orbifold of $\tilde{R}$, even if $H$ is not of the form $H_1 \times
H_2$ for any subgroups $H_1$ of $G_1$ and $H_2$ of $G_2$.

As we have seen in the previous section, in the case of Landau-Ginzburg
orbifolds it is sufficient to compare the $\tau \rightarrow i \infty$ limits of
elliptic genera, \ie\ the Poincar\'{e} polynomials, to establish their
equality for all $\tau$.
With the $\tau \rightarrow i \infty$ limit of $Z[R](g,
g^\prime)$ given by~\efourtw, the corresponding limit of $\hat{Z}[R](g,
\tilde{g})$ is
\eqn\ephat{
\hat{P}[R](g, \tilde{g}) = P^{\rm tw}[R](g) \frac{1}{|G|} \sum_{g^\prime \in G}
\tilde{g}(g^\prime) \prod_{R_i(g) = 1} P^{\rm inv}[R_i](g^\prime).
}
Similarly, we have
\eqn\eptihat{
\hat{P}[\tilde{R}](\tilde{g}, g) = P^{\rm tw}[\tilde{R}](\tilde{g})
\frac{1}{|G|} \sum_{\tilde{g}^\prime \in G^*} \tilde{g}^\prime(g)
\prod_{\tilde{R}_i(\tilde{g}) = 1} P^{\rm
inv}[\tilde{R}_i](\tilde{g}^\prime).
}
Our condition for mirror symmetry now reads
\eqn\enfivethree{
\hat{P}[R](g, \tilde{g}) = \pm \hat{P}[\tilde{R}](\tilde{g}, g).
}
If this condition is fulfilled, then obviously $P[R/H] = \pm
P[\tilde{R}/\tilde{H}]$ for any subgroup $H$. Our results from the last section
then imply that also~\enfivezero\ is satisfied. The natural way for this to
come about is that also~\enfivetwo\ holds. Although we have no proof, we
strongly believe that this is indeed always the case.

The obvious question is now how we may find a pair of Landau-Ginzburg
orbifolds such that~\enfivethree\ is obeyed. To answer this,
we must first introduce
some more notation. Let $s$ be a subset of the set $N$ that indexes the
fields. We will only be
interested in $s$ such that there is at least one element of $G$ which leaves
untwisted the $X_i$ for $i \in s$ and twists the remaining fields. We denote
the set of such $s$ as $S$, and henceforth we will always assume that $s \in
S$. For $g \in G$ we define $\sigma(g) \in S$ by the condition that $i \in
\sigma(g)$ if and only if $X_i$ is left untwisted by $g$. Next, we introduce
the subgroups $G_s$ of elements of $G$ that leave untwisted the fields $X_i$
for $i \in
s$. The remaining fields may be twisted or untwisted depending on which element
of $G_s$ we choose. The corresponding objects in the conjugate
 model are denoted
as $\tilde{s}$, $\tilde{S}$, $\tilde{\sigma}(\tilde{g})$ and $G^*_{\tilde{s}}$
respectively.

To find a pair of conjugate models,
we assume that there is a one-to-one map $\rho$ from
$S$ to $\tilde{S}$ such that $(-1)^{|s|} = (-1)^{|N|-|\rho(s)|}$ and $G_s
\simeq \widetilde{G^*_{\rho(s)}}$ for $s \in S$. Here, $|s|$ denotes the number
of elements in $s$, and as usual the tilde over $G^*_{\rho(s)}$
denotes the dual group. Furthermore, we demand that
\eqn\efivefive{
\prod_{i \in s} P^{\rm inv}[R_i](g) = \sum_{G^*_{\tilde{s}} \subseteq
\widetilde{G_s}} (-1)^{|N|-|\tilde{s}|}
\sum_{\tilde{g} \in G^*_{\tilde{s}}}
\tilde{g}(g^{-1}) P^{\rm tw}[\tilde{R}](\tilde{g}).
}
Our notation means that the first sum runs over all $\tilde{s} \in \tilde{S}$
such that
$G^*_{\tilde{s}} \subseteq \widetilde{G_s}$. We also postulate the
corresponding relation with the roles of the two models interchanged. At this
point, the conditions that we have imposed may seem rather {\it ad hoc}. Our
main justification is that they cover all cases of mirror symmetry between
Landau-Ginzburg orbifolds that we know of. We have performed some limited
computer searches, which support the hypothesis that this is indeed the general
mechanism for mirror symmetry between Landau-Ginzburg orbifolds.

A short calculation~\rPM\ shows that when these conditions are fulfilled
\eqn\eXXX{
\eqalign{
\hat{P}[R](g, \tilde{g})
 = & P^{\rm tw}[R](g) \, P^{\rm tw}[\tilde{R}](\tilde{g}) \sum_{\tilde{g} \in
G^*_{\tilde{s}} \subseteq \widetilde{G_{\sigma(g)}}} (-1)^{|N|-|\tilde{s}|} \cr
\hat{P}[\tilde{R}](\tilde{g}, g) = & P^{\rm tw}[R](g) \, P^{\rm
tw}[\tilde{R}](\tilde{g}) \sum_{g \in G_s \subseteq
\widetilde{G^*_{\tilde{\sigma}(\tilde{g})}}} (-1)^{|N|-|s|}.
}}
where  the sum in, for example, the  first equation runs
$\tilde s\in\tilde S$
 such that $G^*_{\tilde{\sigma}(\tilde{g})} \subseteq
G^*_{\tilde{s}} \subseteq \widetilde{G_{\sigma(g)}}$ and similarly in the
second equation. It follows then that
\eqn\eXXX{
\hat{P}[R](g, \tilde{g}) = (-1)^{|N|} \hat{P}[\tilde{R}](\tilde{g}, g),
}
which proves~\enfivethree. We see that in this way of
implementing mirror symmetry, contributions from twisted fields in one model
correspond to contributions from untwisted fields in the other model and vice
versa, just as it should be for a mirror pair.

\subsec{Examples}\noindent
Finally, we will briefly discuss two classes of solutions, first proposed
in~\rBH, to the conditions
\efivefive. Both of these can be seen as generalizations of
Landau-Ginzburg analogs of the $(2,2)$
minimal models for which mirror symmetry was first discovered~\rGP.
Let us here once again stress that the $N=2$ minimal models
(including products and/or quotients thereof) are the only
theories for which mirror symmetry has been rigorously proven. In terms of
the Landau-Ginzburg models of Fermat type it has been {\it conjectured}
that orbifolds of a Fermat potential come in mirror pairs. This conjecture is
based on studies of the spectra of Landau-Ginzburg vacua~\rMPR\
and their orbifolds~\rMax\ as well as more detailed investigations
of the moduli space of particular $\hat c=3$ theories~\refs{\rCdGP,\rAGM}.
In particular the work in~\refs{\rBH,\rMax} as well as recent advances
in terms of toric geometry~\rBatdual\ indicate that mirror
symmetry must hold for a much larger class of theories than the minimal
models. As was noted in the previous
section, we may construct new solutions by taking the product of
old ones. We conjecture that by taking products of the models we will describe
in this section, one may in fact construct all solutions to the
conditions~\efivefive.

The first class of models is given by a potential of the form~\rArnold\
\eqn\esixone{
W_1 = X_1^{\al_1} + X_1 X_2^{\al_2} + \ldots + X_{N-1} X_N^{\al_N}.
}
The group of phase symmetries is isomorphic to $G \simeq \ZZ_D$, where $D =
\al_1
\ldots \al_N$. Note the particular case $N=1$ for which the model is of
Fermat type which is the conjectured equivalent of the $A_{\a_1-2}$
minimal model. Another special case is $N=2$ and $\al=2$ which corresponds to
the $D$-series.

The conjugate partner of the potential
$W_1$ is a potential of the same type but
with the order of the exponents reversed~\rBH, \ie\
\eqn\esixtwo{
\tilde{W_1} = \tilde{X}_1^{\tilde{\al}_1} + \tilde{X}_1
\tilde{X}_2^{\tilde{\al}_2} + \ldots + \tilde{X}_{N-1}
\tilde{X}_N^{\tilde{\al}_N},
}
where the exponents are given by
\eqn\eXXX{
\tilde{\al}_i = \al_{N+1-i} \;\;  {\rm for} \;\; 1 \leq i \leq N.
}
The group of phase symmetries is obviously isomorphic to $G^* \simeq \ZZ_D$.

Our second example is in many respects similar to the first one. This class of
models is given by a potential of the form
\eqn\eXXX{
W_2 = X_N X_1^{\al_1} + X_1 X_2^{\al_2} + \ldots + X_{N-1} X_N^{\al_N}.
}
The cyclical nature of this potential makes it natural to take the variable
$i$, which indexes the fields, to be defined modulo $N$, \ie\ $i \in \ZZ_N$.
The
group of phase symmetries is isomorphic to $G \simeq \ZZ_D$, where $D = \al_1
\ldots \al_N + (-1)^{N-1}$.
The conjugate
 partner of this potential is a potential of the same type but with
the order of the exponents $\al_i$ reversed~\rBH, \ie\
\eqn\eXXX{
\tilde{W_2} = \tilde{X}_N \tilde{X}_1^{\tilde{\al}_1} + \tilde{X}_1
\tilde{X}_2^{\tilde{\al}_2} + \ldots + \tilde{X}_{N-1}
\tilde{X}_{N}^{\tilde{\al}_N},
}
where the exponents are given by
\eqn\eXXX{
\tilde{\al}_i = \al_{N+1-i} \;\; {\rm for \;\;} i \in \ZZ_N.
}
The group of phase symmetries is isomorphic to $G^* \simeq \ZZ_D$.

Without going into the details, which are tedious but straightforward~\rPM,
one can
show that~\efivefive\ is fulfilled
for both of the above examples. From the general arguments in the previous
section it now follows that the $H\simeq\ZZ_m$ orbifold and the
$\tilde H\simeq\ZZ_{\tilde m}$ orbifold, with $\tilde m=D/m$, of
the Landau-Ginzburg models with potentials $W_{1,2}$ and $\tilde W_{1,2}$
are mirror partners.

\vskip 5mm

{\bf Acknowledgments}:
The authors acknowledge useful discussions with T.~H\"ubsch and E.~Witten.
P.B. was supported by DOE grant DE-FG02-90ER40542 and
M.H. was supported by the Swedish Natural Science Research Council (NFR).

\vfill \eject

\listrefs

\bye